\documentclass{article}
\usepackage[T1]{fontenc}
\usepackage[utf8]{inputenc}
\usepackage{ismir} %
\usepackage{amsmath,cite,url}
\usepackage{graphicx}
\usepackage{color}
\usepackage{array}
\usepackage[export]{adjustbox}
\usepackage{booktabs}
\usepackage{multirow}
\usepackage{subcaption}
\usepackage{float}
\usepackage{kotex}

\usepackage{microtype}
\title{Gugak-VocalSet: A Multi-Genre Studio-Quality Dataset of Korean Traditional Singing with Sigimsae Annotations}

\multauthor
{Danbinaerin Han$^{\dagger*}$ \hspace{0.7cm} Wonil Kim$^{\dagger\ddagger}$ \hspace{0.7cm} Seah Hong$^\S$ \hspace{0.7cm} Dongjoo Moon$^\ddagger$}
{ {\bfseries Jaeok Lee$^\S$ \hspace{0.7cm} Jongpil Lee$^\ddagger$ \hspace{0.7cm} Chaewon Kim$^\S$ \hspace{0.7cm} Juhan Nam$^\dagger$}\\
  $^\dagger$~KAIST, Daejeon \qquad $^\ddagger$~Neutune, Seoul \qquad $^\S$~National Gugak Center, Seoul\\
{\tt\small \{naerin71, juhan.nam\}@kaist.ac.kr}, \quad {\tt\small \{ianwonilkim, mdju22, jongpillee\}@neutune.com} \\
  {\tt\small \{seah0330, jian1223, hilary\}@korea.kr}
}

\def\authorname{D. Han, W. Kim, S. Hong, D. Moon, J. Lee, J. Lee, C. Kim, and J. Nam}

\usepackage[bookmarks=false,pdfauthor={\authorname},pdfsubject={\pdfsubject},hidelinks]{hyperref}

\urldef{\demourl}\url{https://www.danbinaerin.com/Gugak-VocalSet/}
\urldef{\repourl}\url{https://github.com/danbinaerinHan/Gugak-VocalSet}

\begin{document}

\maketitle

\begingroup
\renewcommand{\thefootnote}{\fnsymbol{footnote}}
\footnotetext[1]{The author conducted this work as an intern at Neutune.}
\endgroup

\begin{abstract}
We present Gugak-VocalSet, the first publicly available audio dataset for Korean traditional singing with dense annotations of \emph{sigimsae}, a structured pitch contour system of vocal expressions. Gugak-VocalSet contains 24.5 hours of unaccompanied solo vocal recordings across 305 tracks of representative repertoire curated by the National Gugak Center, performed by 20 professional singers in a controlled studio environment with fixed key and tempo. In addition to genre, lyrics, and mood/timbre tags, the dataset provides a unified sigimsae ontology and approximately 68,000 segments (71,700 instances including multi-label) of audio-grounded sigimsae annotations; the ontology adopts a functional-position-based hierarchy designed for consistent application across genres, and 14 expert annotators specializing in Korean traditional music performed segment-level labeling accordingly.
Baseline experiments on sigimsae classification and event detection show that domain-informed F0 representations remain competitive with, and on classification surpass, frozen probing of a general-purpose music foundation model, while pitch and spectral cues play complementary roles in event-level detection.\footnote{The dataset, experiment code, audio examples, an interactive sigimsae annotation browser, and the full ontology with per-type examples are available through the companion page: \demourl}
\end{abstract}
\section{Introduction}\label{sec:introduction}

In many vocal traditions, pitch expression---continuous pitch movement within and between notes---is not decorative but constitutive of the melodic line. Its realization varies across traditions, all of which has long been a central object of computational music studies~\cite{serra2011multicultural, bozkurt2014computational, rao2014overview, panteli2018review}.
Research in this area has therefore centered on analyzing continuous pitch trajectories derived from audio and annotation vocabularies based on each tradition’s own expressive categories~\cite{gong2016pitch, viraraghavan2017statistical, ranjani2019compact}.

In Korean traditional music, this constitutive role of pitch expression has long been foregrounded under its own name~\cite{Lee1999Sigimsae}. \emph{Sigimsae} encompasses continuous pitch inflections shaping a main note, the dependent ornamental notes attached to it, and transitional gestures between notes. Together with rhythmic cycle (\emph{jangdan}) and melodic line (\emph{garak}), it has been regarded as one of the three fundamental elements of Korean music~\cite{kim2009folk_sigimsae}. Rather than a decorative layer applied to melody, it functions as a structural feature that shapes melodic identity itself. This is why regional melodic systems—\emph{tori} and their melodic paths (\emph{gil})—cannot be reduced to their scales alone; two pieces drawing on the same scale may fall under different tori once their sigimsae part ways~\cite{bohyung2012jo, kim2015gugak}.

When sigimsae is transcribed into staff notation, the continuous audio phenomena must be reduced to symbolic approximations (grace notes, specific ornament symbols), whose form depends on the individual transcriber's interpretation. 
However, the range and variability of sigimsae realization cannot be fully captured by any notation system; any rigorous analysis of actual performance therefore requires working directly from the audio signal rather than from notation-mediated representations.

Despite this need for audio-based analysis, computational research on Korean traditional vocal music has faced two persistent obstacles~\cite{injong2023gugak}. First, publicly available audio resources suitable for research are virtually nonexistent. Although archive construction has long been discussed as an important agenda in Korean music research~\cite{kwon2008TM, Myunghyeon2014NGC, yeri2017archive}, most high-quality recordings are commercially distributed and restricted by copyright,
 while public archive materials consist of uncontrolled field recordings or accompanied performances unsuitable for solo vocal analysis. Second, sigimsae has yet to be systematically annotated at the audio level. Existing discourse on sigimsae has been mediated primarily through staff notation or descriptive text, and terminology and classification vary across genres and lineages, leaving no common ground for cross-genre annotation.
 
This paper presents \textbf{Gugak-VocalSet} to address this gap. The dataset is a corpus of 305 tracks (24.5 hours) of unaccompanied solo vocals across
Korean traditional music genres, recorded in a controlled studio environment. It includes a unified sigimsae ontology based on its position within a note and approximately 68,000 audio-based sigimsae annotations. 
We further provide baselines for sigimsae classification and event detection: a task-specific F0CNN with domain-informed pitch input surpasses a frozen Western-music pre-trained foundation model (MERT) on classification; on event detection, F0 captures ornament identity while mel captures event boundaries, and a lightweight EDTCN remains competitive with MERT.

\section{Related Work}\label{sec:related_work}

\subsection{Non-Western Music Datasets}

Dedicated datasets for non-Western music traditions have been constructed through several systematic efforts.
The CompMusic project initiated systematic corpus construction for Indian, Turkish, Arab, and Chinese music traditions~\cite{serra2014compmusic}, followed by dedicated datasets such as Saraga~\cite{srinivasamurthy2021saraga} for Indian art music, Lyra~\cite{papaioannou2022dataset} for Greek traditional music, IRMA~\cite{irma2025dlfm} for Iranian classical music, and ChMusic~\cite{gong2021chmusic} for Chinese traditional music.
Vocal-centric datasets include Erkomaishvili for Georgian polyphonic singing~\cite{rosenzweig2020erkomaishvili}, VocalNotes for cross-cultural vocal analysis~\cite{vocalnotes2024}, and SANIDHA for studio-quality Carnatic vocal music~\cite{sanidha2025ismir}.

Despite this growing landscape, no dataset targeting Korean traditional vocal music has been introduced.
Two Korean music corpora exist but with limited utility: a nationally funded gugak dataset~\cite{nia2022gugak} is accessible only within Korea and its sigimsae labels comprise only six selected categories. The \textit{Finding Tori} dataset~\cite{han2023finding} covers only a single genre (\emph{tosogminyo}) and does not directly distribute audio.
Gugak-VocalSet fills this gap with openly available multi-genre recordings of Korean traditional vocal music.

\subsection{Vocal Technique and Expressive Pitch Analysis in MIR}

Computational approaches to continuous pitch variation in vocal music have developed across multiple traditions in parallel.
VocalSet~\cite{wilkins2018vocalset} and GTSinger~\cite{zhang2024gtsinger} established broad catalogues of singing techniques,
demonstrating that diverse vocal phenomena can be systematically collected and classified, though both employ file- or clip-level labels.
Event-level work has brought temporal resolution to this effort: timestamp-based detection of nine techniques across 168 J-POP songs~\cite{yamamoto2022analysis, yamamoto2023primadnn}, and the ROD dataset with expert event annotations for six Hindustani \emph{gamaka} types~\cite{kumar2025recognizing}, each addressing the specific vocabulary of its tradition.
Recent work on Indian classical music takes this further: \cite{nuttall2024svaraforms} shows that the same \emph{svara} takes systematically distinct realized forms depending on melodic context, and \cite{bhake2025melodic} quantifies performer-level expressive variation in timing and pitch across repetitions---evidence that pitch inflection is not incidental but governed by tradition-internal rules.
Gugak-VocalSet extends this line of work to Korean traditional vocal music, contributing a cross-genre sigimsae ontology and large-scale event-level annotations for systematic analysis of its expressive pitch system.

\section{Gugak-VocalSet}\label{sec:dataset}
\subsection{Korean Traditional Vocal Music}\label{sec:ktv}

The most widely accepted division of Korean traditional music is between \emph{jeongak}, cultivated by the court and literati, and \emph{minsogak}, formed and transmitted among the common people.
Jeongak and minsogak differ not only in their social foundations but also in musical style, exhibiting distinct aesthetic characteristics in rhythmic cycle management, melodic progression, vocal production, and expressive techniques.
Another important axis of Korean traditional music is regionality. Many repertoires have developed through oral tradition and region- or school-centered transmission systems, resulting in pronounced differences in scale construction, singing method, timbre, and sigimsae usage.
For example, pansori and Namdo folk songs share common scalar characteristics and vocal styles because they originated in the southern region of the Korean peninsula.

\subsection{Data Configuration}\label{subsec:data_config}

\begin{figure}[t]
  \centering
    \includegraphics[width=\columnwidth]{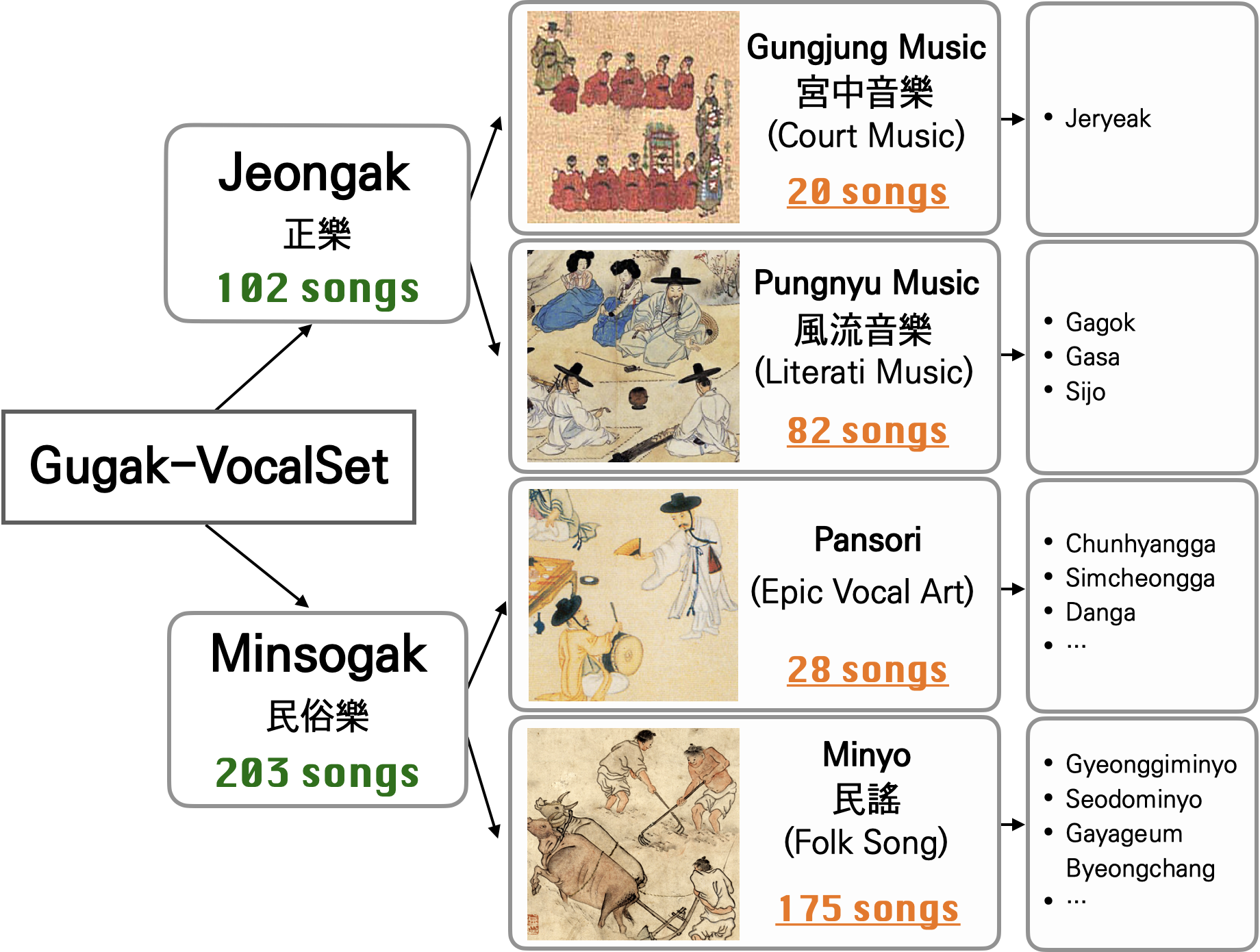}
  \caption{
  Genre hierarchy of Gugak-VocalSet. Numbers show track counts per category.
  }
  \label{fig:genre}
\end{figure}

Gugak-VocalSet is a vocal-centric dataset spanning diverse repertoires of Korean traditional vocal music.
The dataset consists of 305 tracks (approximately 24.5 hours) of representative repertoire curated by the Instrument Research Division of the National Gugak Center (NGC).
The genre hierarchy and the number of pieces in each category are shown in Figure~\ref{fig:genre}.
The sub-genre field encodes genre-specific information: for pansori, it identifies the specific narrative (e.g., \emph{Simcheongga}, \emph{Chunhyangga}); for folk songs, it indicates the representative regional origin.
The pansori tracks were recorded by singers trained in the byeongchang tradition—pansori excerpts with gayageum self-accompaniment—and reflect that lineage's vocal style rather than full-length pansori.

Twenty professional singers participated, each assigned 6–27 tracks, with a mean track duration of 286.7 s (median 260.1 s). The dataset contains 242 female (79.3\%) and 63 male (20.7\%) tracks, with all male tracks belonging to jeongak.

\subsection{Audio Recording \& Refinement}\label{subsec:recording}

All audio was recorded as unaccompanied solo vocals by 20 professional singers in a professional recording studio. Each singer was pre-assigned pieces matching their expertise, with target duration of 4–5 minutes per piece; for shorter pieces, 2–3 lyric variants were recorded within a single track to reach this duration.
Although traditional Korean vocal performance naturally involves flexible tempo and tonic adjustment, each piece was recorded with fixed tonic and tempo to enable corpus-level comparative analysis.
The agreed-upon reference pitch (\emph{cheong}) and a metronome were provided during recording.
Each piece was recorded in two takes; the more stable take was selected by an audio engineer with over 20 years of experience in Korean traditional music recording, and underwent mixing, pitch tuning, tempo correction, and a two-stage quality review.
Both the original and refined audio are provided (WAV, 48\,kHz, 24-bit dual-mono).

\subsection{Annotations \& Metadata}\label{subsec:metadata}
Gugak-VocalSet provides multi-layered metadata and annotations alongside the audio, summarized in Table~\ref{tab:annotations}.

\begin{table}[t]
\centering
\small
\begin{tabular}{@{}p{1.0cm}p{1.5cm}p{4.5cm}@{}}
\toprule
\textbf{Level} & \textbf{Type} & \textbf{Content} \\
\midrule
\multirow{4}{*}{Song}
& \multirow{2}{*}{Metadata} & Title, Genre (top/mid/sub), Singer, Jangdan, Tempo, Time Sig., EN/KR Captions \\
\cmidrule(l){2-3}
& \multirow{2}{*}{Expert Tags} & Key, Scale, Mood (80 tags), Timbre (38 tags) \\
\midrule
\multirow{2}{*}{Segment} & Lyrics & Phrase-level (7--20\,s) \\
& Alignment & Lyrics--Audio Alignment \\
\midrule
Event & Sigimsae & 17 Ornament Types w/ Timestamps \\
\bottomrule
\end{tabular}
\caption{Summary of metadata and annotations in Gugak-VocalSet.}
\label{tab:annotations}
\end{table}

For each track, we provide basic metadata including genre labels, singer name and gender, jangdan, tempo, and a Western key label. The Western key is assigned by Western-music-trained annotators through subjective listening as the tonality most compatible with each piece, and is provided to support general-purpose applications. The basic metadata also includes bilingual (Korean/English) captions and historical/musical context drawn from the ``Gugak Dictionary'' published by the NGC. In addition, we provide expert-curated mood (80 categories) and timbre (38 categories) tags, annotated by five domain experts in Korean traditional music.

\begin{figure*}[t]
  \centering
  \includegraphics[width=0.9\textwidth]{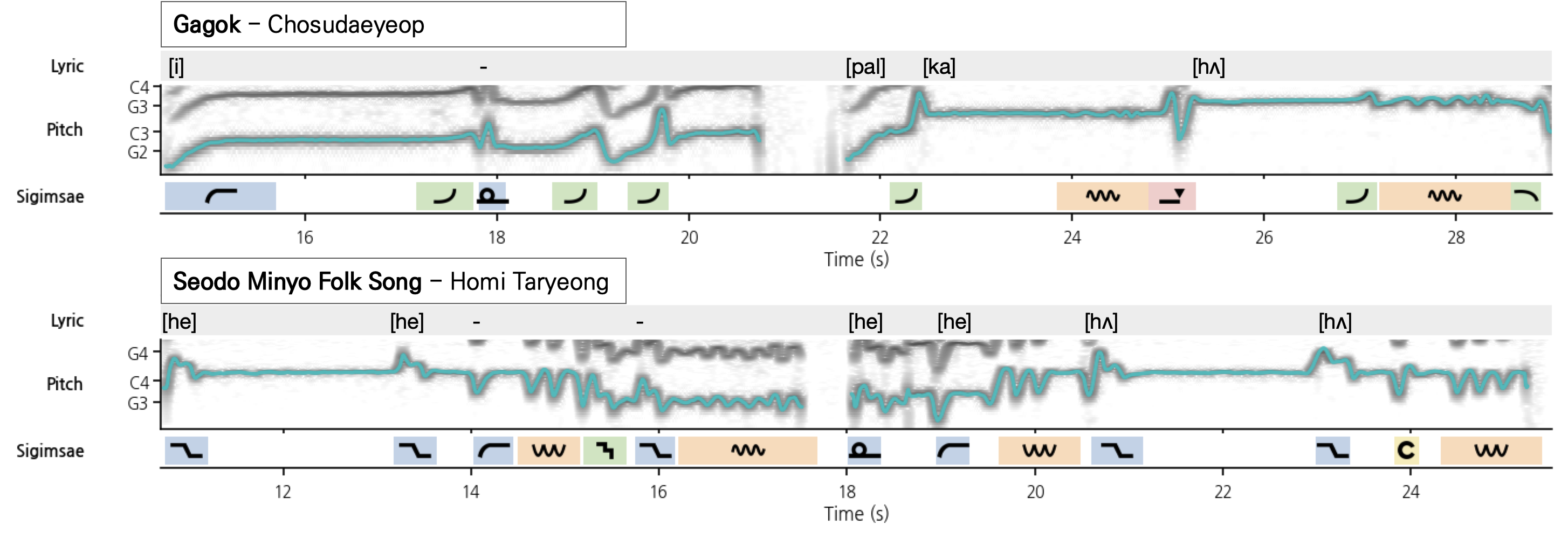}
  \caption{Annotation examples: \textit{jeongak} (top) and \textit{minsogak} (bottom). Each panel shows lyric--audio alignment, F0 contour overlaid on a spectrogram, and sigimsae annotations color-coded by functional category.}
  \label{fig:annotation}
\end{figure*}

Lyrics--audio alignment was performed at the phrase level (7--20\,s) according to jangdan structure, transcribed from the text actually sung by each performer. This alignment and the sigimsae annotations share the same annotator team and workflow, detailed in Section~\ref{subsec:annotation_process}. Figure~\ref{fig:annotation} shows the resulting three-layer annotation---aligned lyrics, F0 contour, and sigimsae regions---for a jeongak and a minsogak track.

\begin{figure}[!t]
\centering
\scriptsize
\setlength{\tabcolsep}{1.5pt}
\renewcommand{\arraystretch}{1.0}
\begin{tabular}{@{}p{2.3cm}@{\hskip 3pt}m{0.8cm}@{\hskip 0.5pt}m{1.0cm}@{\hskip 2pt}p{3.4cm}@{}}
\toprule
\textbf{Name} & \textbf{Sym.} & \textbf{Contour} & \textbf{Description} \\
\midrule
\multicolumn{4}{l}{\scriptsize\textbf{Onset Sigimsae (앞시김새)}} \\[1pt]
onset asc.\ slide\newline{\tiny(밀어내기)}
 & \includegraphics[height=0.3cm,valign=c]{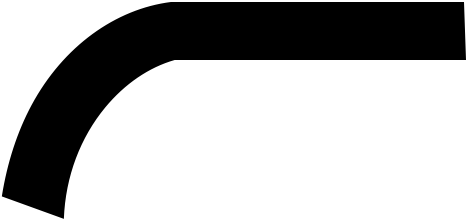}
 & \includegraphics[height=0.6cm,valign=c]{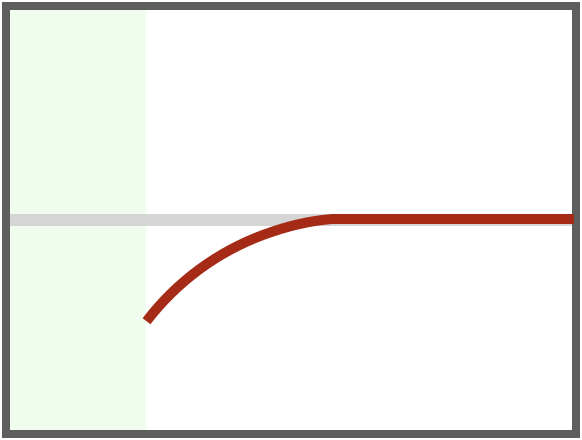}
 & Sliding up from a lower pitch to the main note \\\addlinespace[2pt]
onset desc.\ flick\newline{\tiny(꺾어내기)}
 & \includegraphics[height=0.3cm,valign=c]{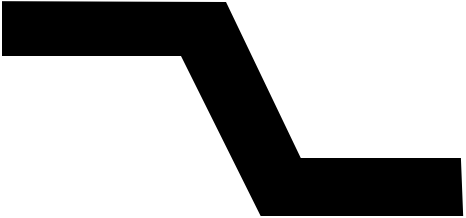}
 & \includegraphics[height=0.6cm,valign=c]{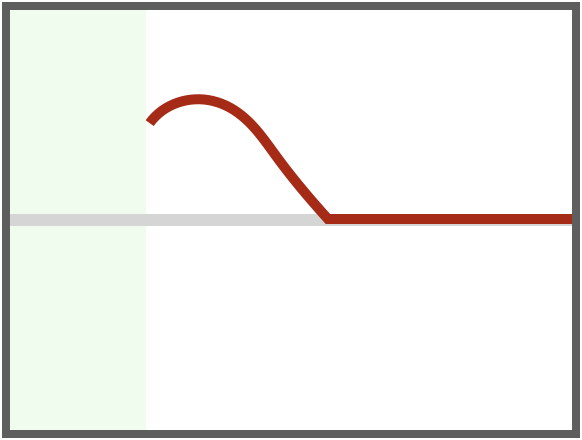}
 & Briefly touching a higher pitch before flicking down to the main note \\\addlinespace[2pt]
onset upper turn\newline{\tiny(감아내기 위)}
 & \includegraphics[height=0.3cm,valign=c]{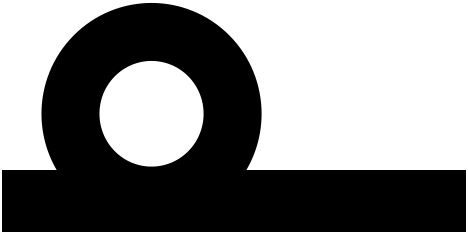}
 & \includegraphics[height=0.6cm,valign=c]{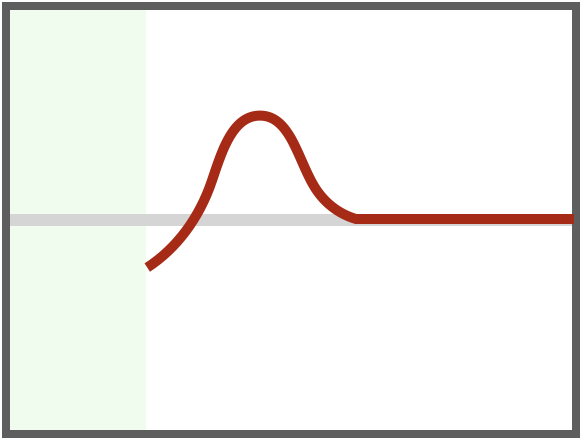}
 & Winding upward, passing through a higher pitch, then returning to the main note \\\addlinespace[2pt]
onset lower turn\newline{\tiny(감아내기 아래)}
 & \includegraphics[height=0.3cm,valign=c]{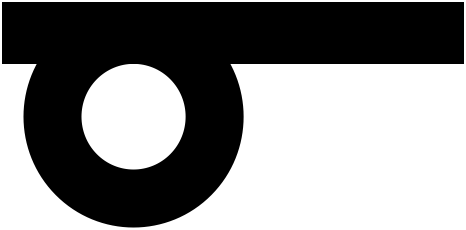}
 & \includegraphics[height=0.6cm,valign=c]{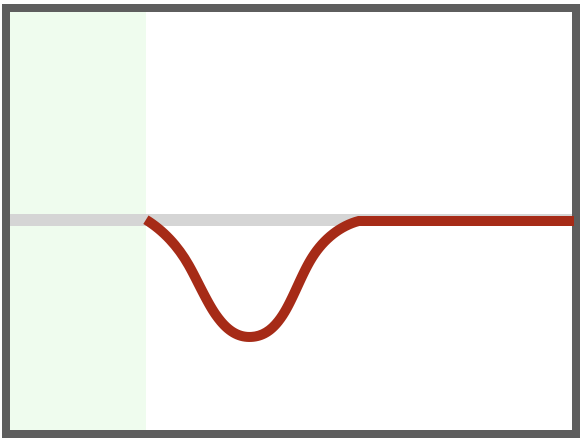}
 & Winding downward, passing through a lower pitch, then returning to the main note \\
\midrule
\multicolumn{4}{l}{\scriptsize\textbf{Vibrato Sigimsae / (요성 시김새) / \emph{yoseong}  / 搖聲}} \\[1pt]
standard vibrato\newline{\tiny(보통 요성)}
 & \includegraphics[height=0.24cm,valign=c]{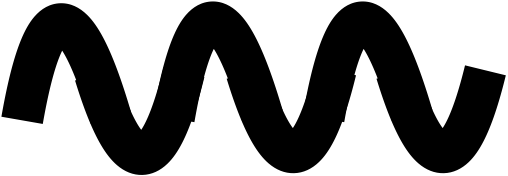}
 & \includegraphics[height=0.6cm,valign=c]{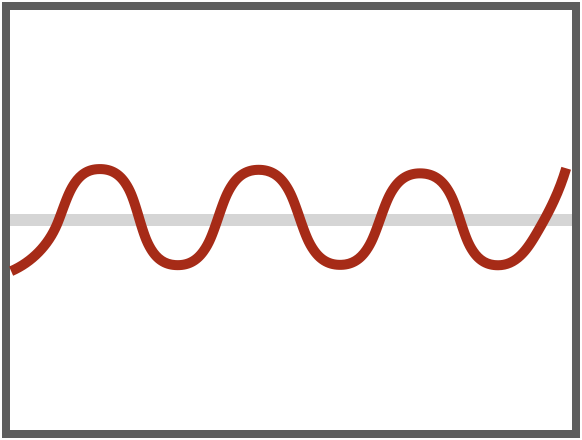}
 & Continuously oscillating above and below the main note ($<$4th) \\\addlinespace[2pt]
wide vibrato\newline{\tiny(격한 요성)}
 & \includegraphics[height=0.34cm,valign=c]{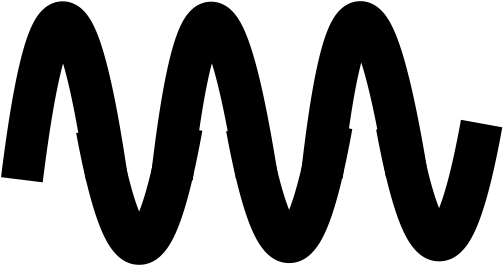}
 & \includegraphics[height=0.6cm,valign=c]{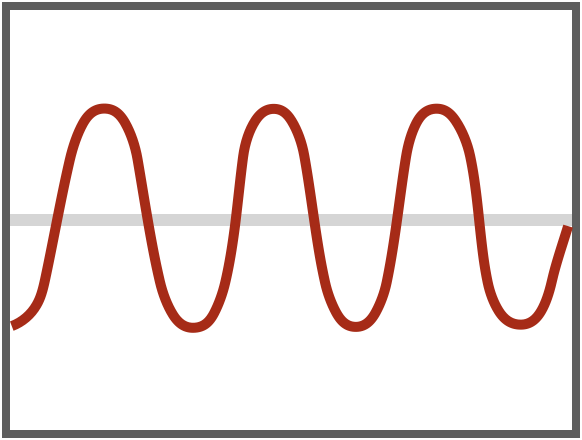}
 & Vigorously oscillating above and below the main note ($\geq$4th) \\\addlinespace[2pt]
lower-cent.\ vibrato\newline{\tiny(아래로 향하는 요성)}
 & \includegraphics[height=0.22cm,valign=c]{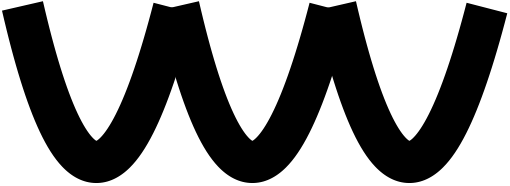}
 & \includegraphics[height=0.6cm,valign=c]{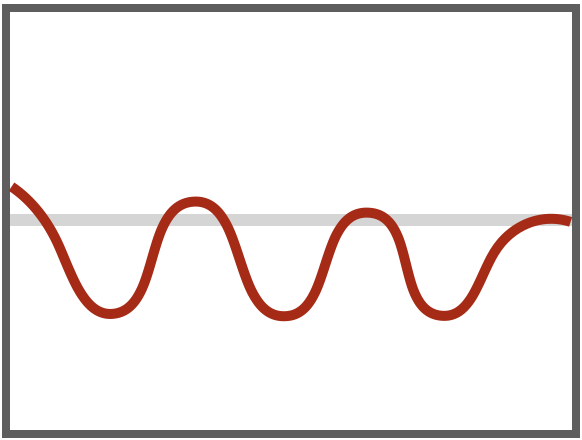}
 & Oscillating predominantly below the main note; settles back on it \\\addlinespace[2pt]
desc.-tail vibrato\newline{\tiny(흘러내리는 요성)}
 & \includegraphics[height=0.37cm,valign=c]{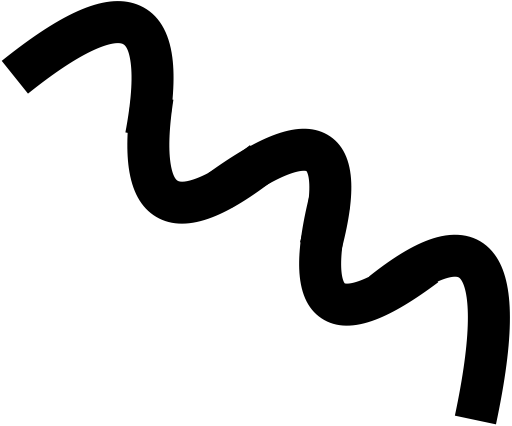}
 & \includegraphics[height=0.6cm,valign=c]{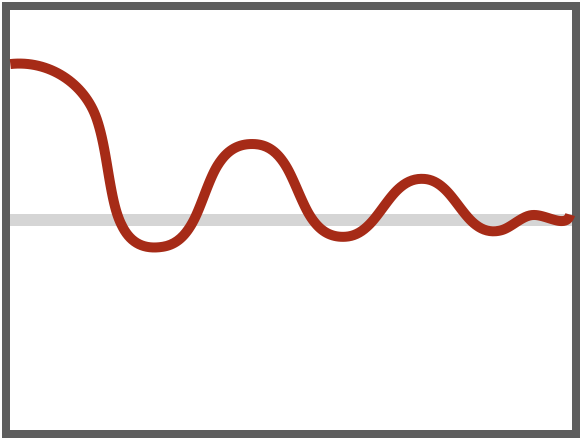}
 & Oscillating while gradually descending \\
\midrule
\multicolumn{4}{l}{\scriptsize\textbf{Mid-note Sigimsae (중간 시김새)}} \\[1pt]
mid-note upper roll\newline{\tiny(굴려내기 위)}
 & \includegraphics[height=0.34cm,valign=c]{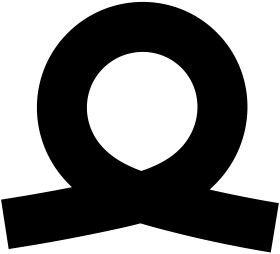}
 & \includegraphics[height=0.6cm,valign=c]{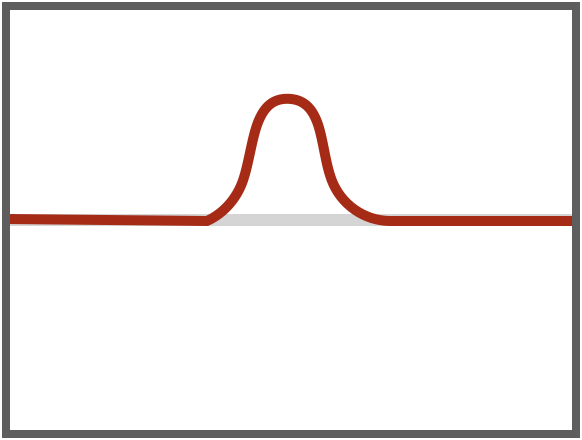}
 & Briefly touching a pitch above the main note as ornament \\\addlinespace[2pt]
mid-note lower roll\newline{\tiny(굴려내기 아래)}
 & \includegraphics[height=0.34cm,valign=c]{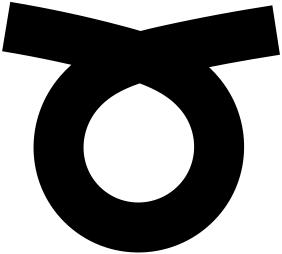}
 & \includegraphics[height=0.6cm,valign=c]{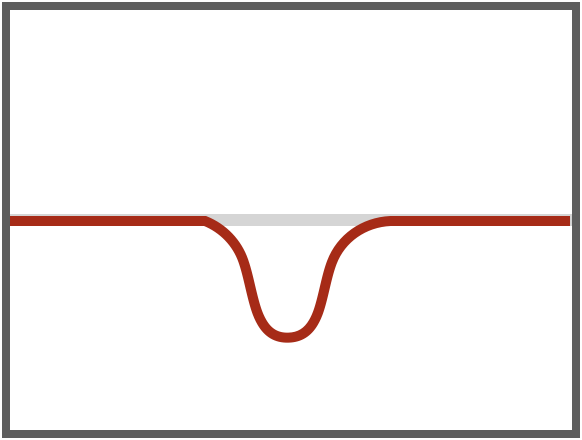}
 & Briefly touching a pitch below the main note as ornament \\
\midrule
\multicolumn{4}{l}{\scriptsize\textbf{Offset Sigimsae (뒷시김새)}} \\[1pt]
offset asc.\ slide\newline{\tiny(추성 / \emph{chuseong}  / 推聲)}
 & \includegraphics[height=0.3cm,valign=c]{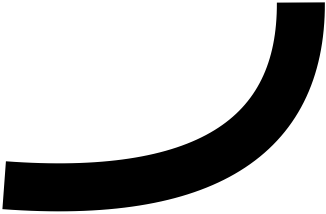}
 & \includegraphics[height=0.6cm,valign=c]{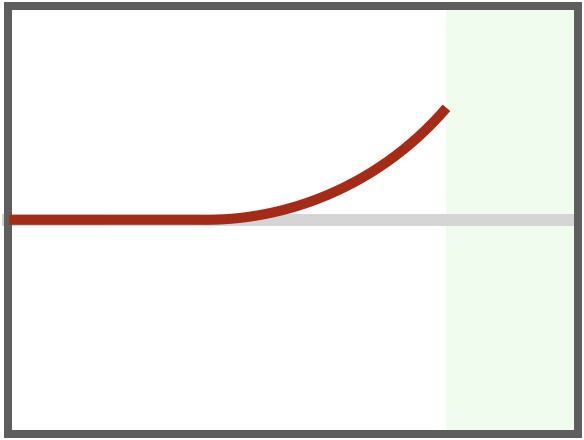}
 & Sliding upward from the main note toward the end \\\addlinespace[2pt]
offset desc.\ slide\newline{\tiny(퇴성 / \emph{toeseong} / 退聲)}
 & \includegraphics[height=0.25cm,valign=c]{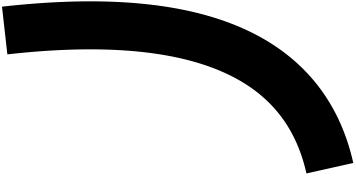}
 & \includegraphics[height=0.6cm,valign=c]{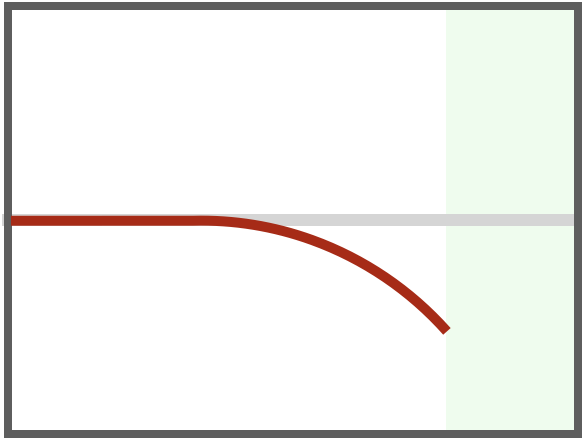}
 & Sliding downward from the main note toward the end \\\addlinespace[2pt]
offset roll-back\newline{\tiny(끝굴려내기 위)}
 & \includegraphics[height=0.34cm,valign=c]{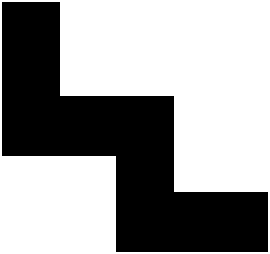}
 & \includegraphics[height=0.6cm,valign=c]{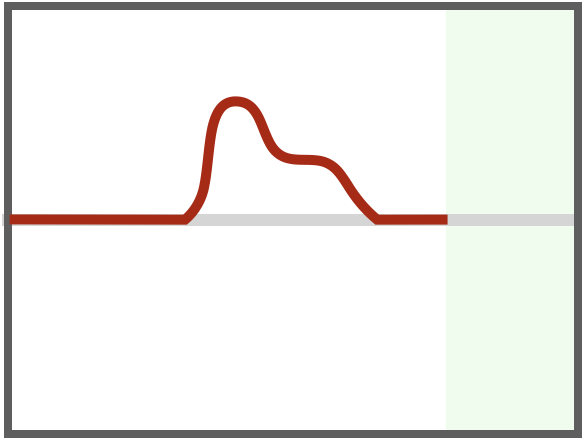}
 & Rolling through pitches above the main note, returning to it \\\addlinespace[2pt]
offset lower roll-away\newline{\tiny(끝굴려내기 아래)}
 & \includegraphics[height=0.34cm,valign=c]{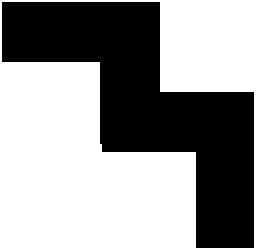}
 & \includegraphics[height=0.6cm,valign=c]{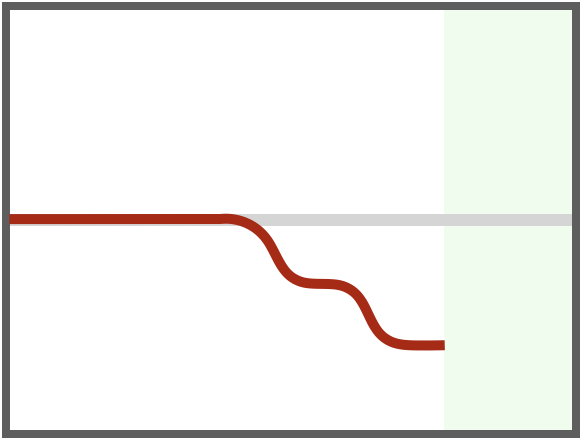}
 & Rolling through pitches below the main note, descending away \\\addlinespace[2pt]
offset sharp cut\newline{\tiny(끊어내기)}
 & \includegraphics[height=0.3cm,valign=c]{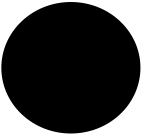}
 & \includegraphics[height=0.6cm,valign=c]{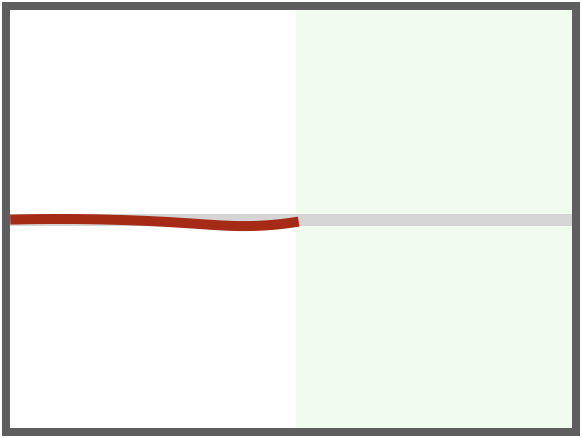}
 & Abruptly cutting off the tone \\
\midrule
\multicolumn{4}{l}{\scriptsize\textbf{Accent Sigimsae (강조 시김새)}} \\[1pt]
upward accent\newline{\tiny(전성 / \emph{jeonseong}  / 轉聲)}
 & \includegraphics[height=0.3cm,valign=c]{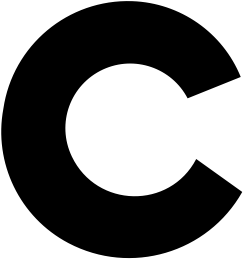}
 & \includegraphics[height=0.6cm,valign=c]{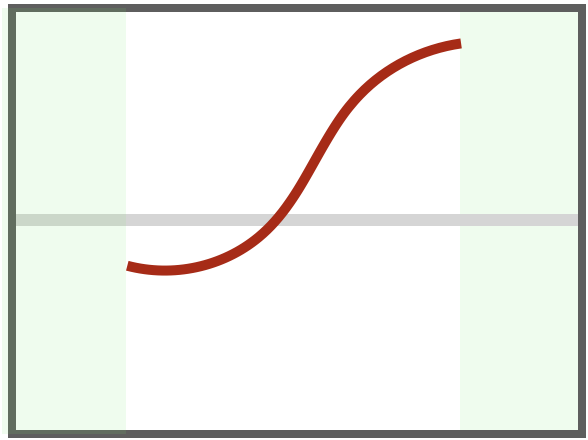}
 & Rapidly striking upward to accentuate the main note \\
\midrule
\multicolumn{4}{l}{\scriptsize\textbf{Transition Sigimsae (경과 시김새)}} \\[1pt]
transition upward snap\newline{\tiny(채올리기)}
 & \includegraphics[height=0.3cm,valign=c]{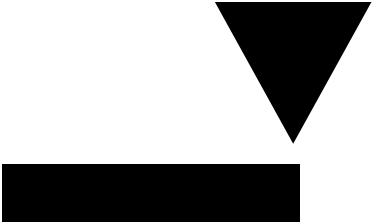}
 & \includegraphics[height=0.6cm,valign=c]{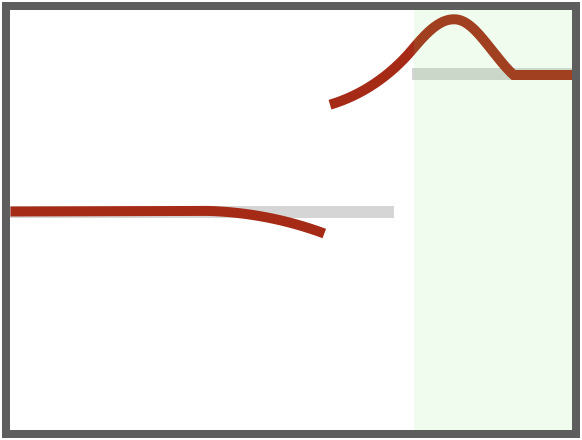}
 & Briefly snapping up above the following note as transition \\
\bottomrule
\end{tabular}
\caption{The 17 sigimsae types in the unified ontology grouped by functional position, with notation symbol, pitch contour, and description.}
\label{fig:sigimsae_ontology}
\end{figure}

\section{Sigimsae Annotation}\label{sec:sigimsae}
\subsection{Unified Sigimsae Ontology}\label{subsec:ontology}

Existing discourse on sigimsae has been conducted primarily through staff notation or descriptive text, and terminology and classification systems vary across genres and transmission lineages. The same concept may bear different names in different genres, or distinct concepts may be conflated under similar terms; attributes of different dimensions---performance technique, ornamental function, and perceptual impression---are sometimes treated at the same level. Due to these inconsistencies, no audio-based cross-genre sigimsae annotation system has been established.

To address this, we designed a unified ontology applicable consistently regardless of genre or transmission method.
The design proceeded in two stages: first, sigimsae notations recorded in \textit{jeongga} and \textit{pansori} score collections were surveyed to identify types distinguishable by pitch-contour shape; a prototype annotation round then revealed additional genre-specific types that had been missed, leading to the final set of 17 types.
Core distinctions traditionally recognized across genres were preserved, while genre-specific vocal production and timbre distinctions were excluded in favor of focusing on pitch variation and note-transition patterns. On this basis, functional position within the melodic line---onset, mid-note, offset, and transition---serves as the top-level hierarchy, organizing 17 sigimsae types (Figure~\ref{fig:sigimsae_ontology}).

For naming, established traditional terms were retained where unambiguous (\emph{chuseong}, \emph{toeseong}, \emph{jeonseong}, \emph{yoseong}, etc.), while other types were given intuitive Korean names reflecting their contour shape and function. Additionally, a symbol system was introduced that visually reflects pitch-contour shapes in conventional notation symbols, so that the acoustic characteristics of each type are intuitively apparent from the symbol (Figure~\ref{fig:sigimsae_ontology}).

\subsection{Annotation Process and Statistics}\label{subsec:annotation_process}

The sigimsae annotation and the phrase-level lyrics--audio alignment (Section~\ref{subsec:metadata}) were performed by the same team of 14 annotators specializing in Korean traditional vocal music; the project member who designed the sigimsae ontology (Section~\ref{subsec:ontology}) and is trained in Korean traditional music performed the label-correction pass. Annotators were assigned pieces according to their specialization genre; for genres without direct specialists, those familiar with the most similar vocal technique were assigned to ensure familiarity with the relevant sigimsae conventions. Prior to annotation, guidelines and reference pages containing audio examples and pitch-contour visualizations for each sigimsae type were provided.

Annotations underwent a two-stage review process. In the first stage, error patterns commonly observed across all annotators were identified, and correction guidelines were communicated to annotators with a request for re-annotation. In the second stage, the reviewer examined the revised results and corrected remaining problematic segments to finalize the annotations.
Multi-label annotation was permitted, but annotators were instructed to segment regions into single sigimsae types wherever possible.

\begin{figure}[t]
  \centering
  \begin{subfigure}[b]{\columnwidth}
    \centering
    \includegraphics[width=\columnwidth]{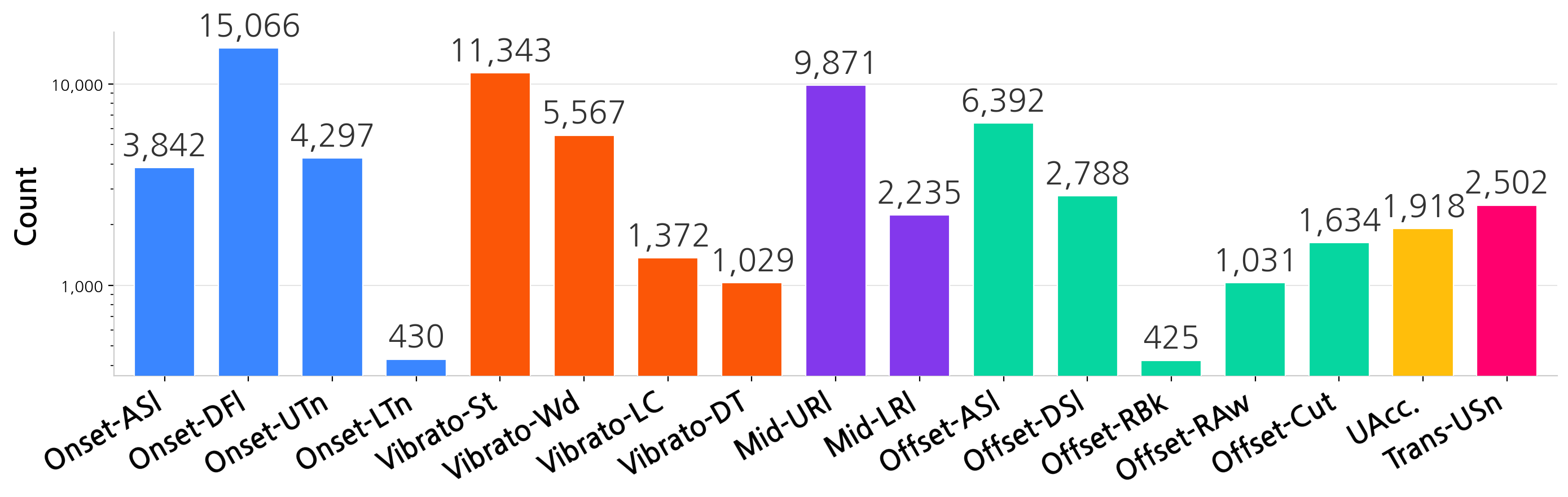}
    \caption{Annotation count per type (log scale).}
    \label{fig:sigimsae_count}
  \end{subfigure}
  \vspace{0.3em}
  \begin{subfigure}[b]{\columnwidth}
    \centering
    \includegraphics[width=\columnwidth]{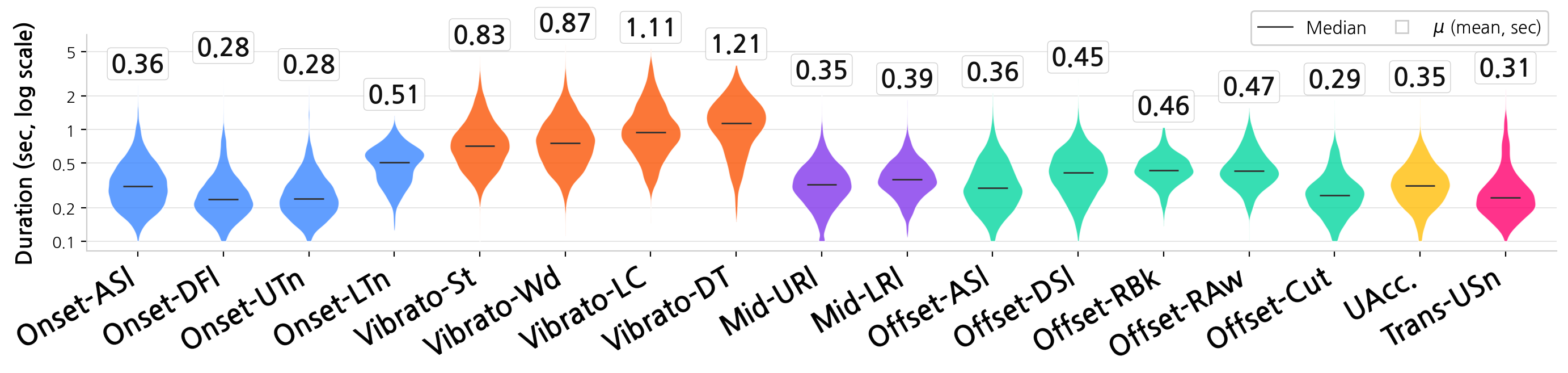}
    \caption{Duration distribution per type (violin plot, log scale).}
    \label{fig:sigimsae_dur}
  \end{subfigure}
  \caption{Statistics of 71,742 sigimsae instances across 305 tracks. (a)~Annotation count per type (log scale). (b)~Duration distribution per type; labels show means $\mu$.}
  \label{fig:sigimsae_stats}
\end{figure}

A total of 71,742 sigimsae instances were annotated across 68,272 segments in 305 tracks, of which 3,453 segments (5.1\%) carry multi-label annotations. The overall mean segment duration is 0.47\,s (median 0.35\,s, IQR [0.23, 0.57]\,s).
Figure~\ref{fig:sigimsae_stats}(a) shows the annotation count per type and
Figure~\ref{fig:sigimsae_stats}(b) presents the duration distribution of each type as violin plots. Vibrato types exhibit notably longer durations (median 0.76\,s) compared to other ornaments, reflecting the sustained oscillatory nature of vibrato.

We quantify the annotation outcome along two axes: review impact and granularity. Throughout, we use event-level F1 with a 200\,ms collar (EvF1@200ms), matching two events when onset and offset each agree within 200\,ms; \emph{class-agnostic} scores require only such a temporal match, while \emph{class-aware} scores additionally require the same 17-class label. \textbf{(i) Scope of post-hoc revision.} Comparing pre- and post-review labels on the tracks subjected to the reviewer's correction pass yields EvF1@200ms $=$ .903 and onset F1 $=$ .910, with 93.9\% of time-agreed events retaining the same 17-class label, and total region count varying by only 1.5\% before and after review. The review thus provided targeted corrections, and the released labels largely reflect the work of the 14 specialist annotators.
\textbf{(ii) Granularity analysis. }Given the single-specialist-per-piece protocol described above, we conducted a focused consistency check: the ontology designer and a second project member each independently re-annotated the first 60\,s of 10 tracks (10\,min total). Between their annotations, class-agnostic EvF1@200ms reaches .597 (onset F1 .720), while class-aware EvF1@200ms drops to .357: the two annotators largely agree on \emph{where} sigimsae events occur, but often diverge on \emph{which} of the 17 fine-grained types to assign. Consistent with this, merging the most-confused 17-class pairs into a 10-class coarse grouping raises the same-label rate from 56\% to 78\%, indicating that fine-level disagreement is concentrated in a small number of acoustically adjacent pairs rather than distributed across the ontology.

\section{Baseline Experiments}\label{sec:experiments}

Building on these annotations, we evaluate sigimsae classification and event detection on the full 17-category ontology.

\subsection{Experimental Setup}\label{subsec:setup}
We apply song-level splits to prevent track leakage between train and test. Multi-label segments (5.1\%) are resolved by primary-label reduction for yoseong-dominant cases and skipped for other ambiguous combinations; full multi-label handling is left for future work.
We extract F0 with RMVPE~\cite{wei2023rmvpe} at 10\,ms hop, represented as 3 channels: cents-normalized contour (segment-median reference for classification, song-median for detection), binary voicing, and $\Delta$F0 (scaled for detection). Mel spectrograms are computed at 22.05\,kHz with 128 bins, 2048-sample windows, and 512-sample hop ($\approx$23.2\,ms). MERT-family models consume raw audio at 24\,kHz with their native 320-sample hop ($\approx$13.3\,ms).

\subsection{Task 1: Sigimsae Classification}\label{subsec:classification}

The sigimsae classification task requires predicting the type of a pre-segmented sigimsae region (mean duration 0.47\,s), thereby testing the acoustic discriminability across the 17 ontology categories. We design baselines along two axes: input representation (F0 contour vs.\ Mel spectrogram) for task-specific CNNs, and frozen music foundation models---MERT-v1~\cite{li2023mert} and its culturally-adapted variants CultureMERT and CultureMERT-TA~\cite{kanatas2025culturemert}---probed with a MARBLE-style 2-layer MLP head (hidden=512) over frozen hidden states. 
CultureMERT applies continual pre-training (CPT) to MERT-v1-95M on 650h of non-Western traditions (Turkish makam, Hindustani, Carnatic, Greek Lyra); we include it under the hypothesis that shared traits such as modal structure, microtonal inflection, and ornamented melodic lines may transfer to Korean traditional music, despite Korean music not being part of its CPT corpus.
For each foundation model we probe layers L$\in\{5,7,9,11,12\}$ and report the best layer per model. Probing here measures whether linearly separable features for sigimsae are already present in the pre-trained representation.

\textbf{Evaluation.}
We use song-level 5-fold cross-validation (per-fold 60/20/20 train/val/test split of songs) and report Macro F1 and Weighted F1 as mean$\pm$std over folds.

\begin{table}[t]
\centering
\resizebox{\columnwidth}{!}{%
\begin{tabular}{llcc}
\toprule
Model & Input & Macro F1 & Weighted F1 \\
\midrule
\multicolumn{4}{l}{\textit{Task-specific CNNs (trained from scratch)}} \\
F0CNN   & F0 contour  & \textbf{.441}{\scriptsize$\pm$.015} & \textbf{.530}{\scriptsize$\pm$.022} \\
MelCNN  & Mel spec    & .400{\scriptsize$\pm$.018} & .510{\scriptsize$\pm$.025} \\
\midrule
\multicolumn{4}{l}{\textit{Frozen foundation models (best layer, MARBLE 2-layer MLP probe)}} \\
MERT-v1\textsubscript{95M} (L7)           & MERT emb. & .358{\scriptsize$\pm$.011} & .454{\scriptsize$\pm$.017} \\
CultureMERT-TA\textsubscript{95M} (L7)    & MERT emb. & .334{\scriptsize$\pm$.022} & .427{\scriptsize$\pm$.031} \\
CultureMERT\textsubscript{95M} (L5)       & MERT emb. & .316{\scriptsize$\pm$.017} & .410{\scriptsize$\pm$.019} \\
\bottomrule
\end{tabular}%
}
\caption{17-category sigimsae classification (song-level 5-fold CV, mean±std). MERT-family models: best layer among L$\in\{5,7,9,11,12\}$.}
\label{tab:cls_17cat}
\end{table}

\textbf{Results.}
Table~\ref{tab:cls_17cat} summarizes the 17-category classification results. F0CNN (Macro F1 .441) surpasses MelCNN (.400), suggesting that the F0 contour alone captures most of the discriminative signal for sigimsae and indicating the strength of domain-informed pitch representations.
In contrast, frozen MERT-v1\textsubscript{95M} peaks at Macro F1 .358 at its best probing layer (L7), falling short of the task-specific CNNs. CultureMERT-TA\textsubscript{95M} (L7, .334) and CultureMERT\textsubscript{95M} (L5, .316) underperform MERT-v1\textsubscript{95M} at every probed layer. Across matched 95M backbones, CPT thus incurs a Macro F1 loss of roughly 0.02–0.04. Considering that Korean music is not part of CultureMERT's CPT corpus (Turkish, Indian, Greek), this points to the importance of including the target domain in pre-training.

\textbf{Layer shift.} 
A layer ablation across L$\in\{5,7,9,11,12\}$ reveals a consistent pattern: all three foundation models perform worst at the final layer (L12) and peak at intermediate layers (L5–L7). This is consistent with prior MIR observations that higher-level semantic information such as genre and style is concentrated in the upper layers of MERT-family models, while acoustic and pitch-level ornament information is more distributed across intermediate layers \cite{zhou2025layer}.

\begin{figure}[t]
\centering
\includegraphics[width=\columnwidth]{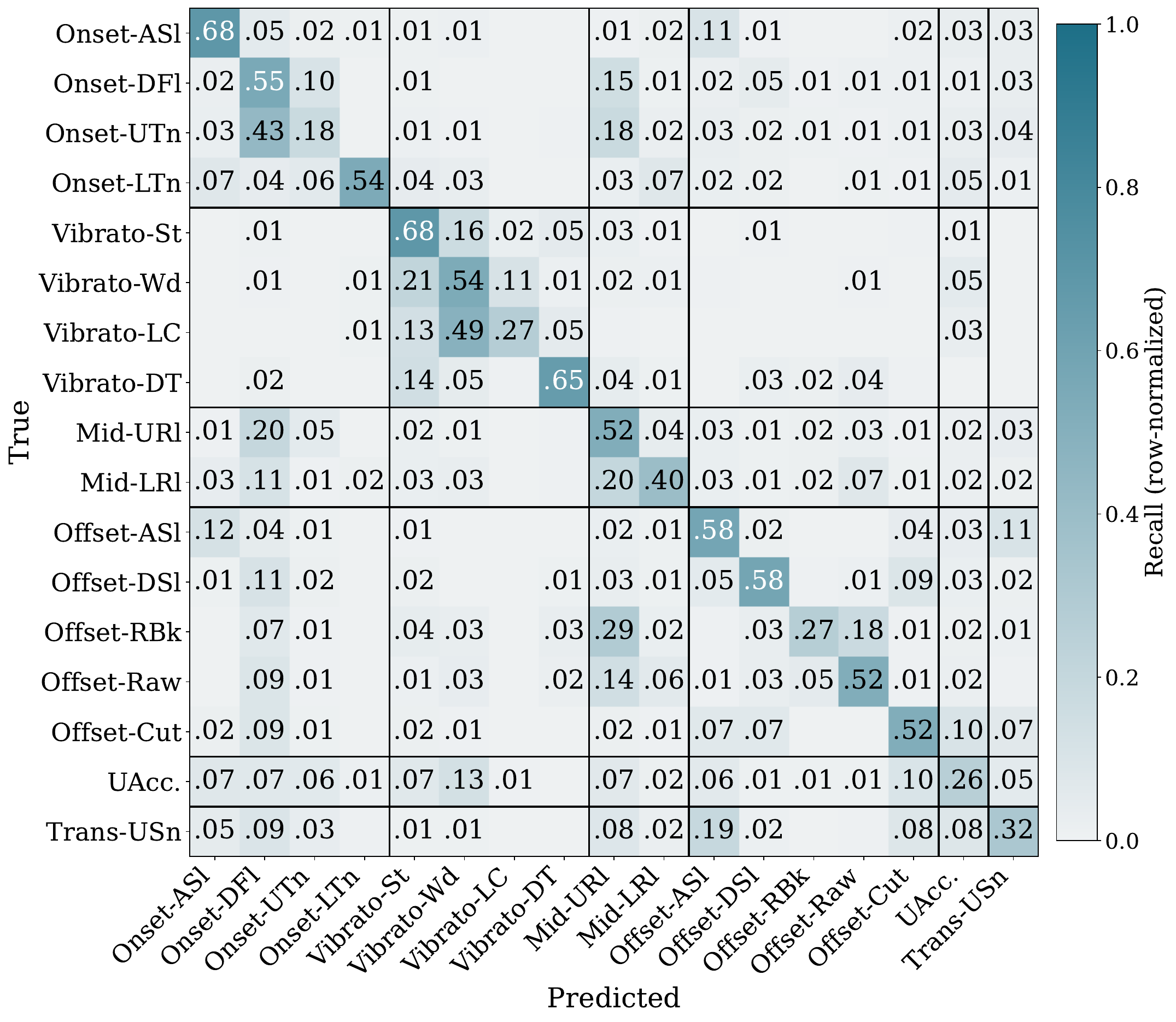}
\caption{Row-normalized (recall) confusion matrix for 17-type classification (F0CNN, aggregated over 5 folds).}
\label{fig:cm_17type}
\end{figure}

\textbf{Error patterns.}
Figure~\ref{fig:cm_17type} shows class-level error patterns for F0CNN. Confusions among yoseong subtypes, and between Onset-DFl and Onset-UTn (both rising-then-falling onset contours) are prominent. 
  A 10-class coarse grouping derived from these model confusions mirrors the IAA-derived grouping (Sec~\ref{subsec:annotation_process}): applied to the IAA pair, both groupings recover the same $\sim$78\% same-label rate (78.4\% vs.\ 78.0\%; 239 vs.\ 238
  of 305) despite independent confusion sources --- evidence that fine-level disagreement reflects acoustic ambiguity rather than annotation noise.

\begin{table}[t]
\centering
\resizebox{\columnwidth}{!}{%
\begin{tabular}{lcccc}
\toprule
 & \multicolumn{2}{c}{Within} & \multicolumn{2}{c}{Cross} \\
\cmidrule(lr){2-3} \cmidrule(lr){4-5}
Direction & M.F1 & Acc & M.F1 & Acc \\
\midrule
J $\rightarrow$ M & .423{\scriptsize$\pm$.007} & .558{\scriptsize$\pm$.005} & .279{\scriptsize$\pm$.002} & .429{\scriptsize$\pm$.010} \\
M $\rightarrow$ J & .354{\scriptsize$\pm$.004} & .500{\scriptsize$\pm$.006} & .286{\scriptsize$\pm$.013} & .432{\scriptsize$\pm$.035} \\
\bottomrule
\end{tabular}%
}
\caption{Cross-genre Macro F1 (F0CNN, 17-cat, 5-fold CV, 102 songs per genre, mean$\pm$std over 3 seeds).}
\label{tab:cross_genre_17cat}
\end{table}

\textbf{Cross-genre generalization.}
We cross-train between jeongak (J) and minsogak (M), matching both at 102 pieces to control for data-size imbalance (Table~\ref{tab:cross_genre_17cat}).
Cross-genre F1 drops 34.0\% for J$\rightarrow$M and 19.2\% for M$\rightarrow$J (cross Acc $\approx$.43), well above chance and indicating a shared acoustic foundation.
The asymmetry may reflect greater variability in minsogak sigimsae, though gender is fully confounded with genre (all male tracks are jeongak), so register differences may also contribute.

\subsection{Task 2: Sigimsae Event Detection}\label{subsec:event_detection}

Sigimsae event detection is a more practical task that jointly predicts the presence, onset/offset boundaries, and type of sigimsae from a continuous audio stream.
Frame-level classification at each representation's native hop (10--23\,ms) outputs per-frame probabilities over 18 classes (17 sigimsae categories + no\_ornament).
We study the effect of input representation on a single EDTCN backbone (encoder--decoder TCN with dilations 1--4; ${\sim}$0.8M params)~\cite{lea2017temporal}, comparing \textbf{F0} and \textbf{mel} (features from Sec.~\ref{subsec:setup}). As self-supervised references we include frozen MERT-v1\textsubscript{95M} and CultureMERT\textsubscript{95M}, each probed with a MERTHead (LayerNorm + 2-layer BiGRU, hidden=128 + linear).
Training employs focal loss ($\gamma$=2.0), don't-care masking~\cite{kumar2025recognizing}, and $\pm$0.5\,s temporal jittering; post-processing applies a median filter (kernel=11 frames) and 50\,ms gap merging.

\textbf{Evaluation.}
Beyond frame-level accuracy and macro F1, we report event-level F1 ($\pm 200\,$ms onset/offset collar) and segment-level F1 (IoU $\geq 0.5$).
\begin{table}[t]
\centering
\resizebox{\columnwidth}{!}{%
\begin{tabular}{lccc}
\toprule
Model & Frame F1\textsubscript{M} & Ev.F1\textsubscript{200ms} & IoU F1\textsubscript{0.5} \\
\midrule
EDTCN (F0)                          & \textbf{.324}{\scriptsize$\pm$.015} & .675{\scriptsize$\pm$.014} & .661{\scriptsize$\pm$.014} \\
EDTCN (mel)                         & .269{\scriptsize$\pm$.013} & \textbf{.763}{\scriptsize$\pm$.012} & \textbf{.712}{\scriptsize$\pm$.015} \\
\midrule
MERT-v1\textsubscript{95M} (L5)     & .311{\scriptsize$\pm$.015} & .756{\scriptsize$\pm$.026} & .706{\scriptsize$\pm$.028} \\
CultureMERT\textsubscript{95M} (L5) & .294{\scriptsize$\pm$.016} & .747{\scriptsize$\pm$.019} & .700{\scriptsize$\pm$.022} \\
\bottomrule
\end{tabular}%
}
\caption{Sigimsae event detection results on the 17-category ontology. MERT-family: best layer in L$\in\{5,7,9,12\}$ by frame Macro F1.}
\label{tab:event_detection}
\end{table}

\textbf{Results.}
Table~\ref{tab:event_detection} reports the event detection results.
Pitch (F0) and spectral (mel) inputs play complementary roles: F0 leads on frame-level Macro F1 (.324 vs.\ .269 for mel and .311 for MERT-v1), while mel leads on event-level metrics. The lightweight EDTCN(mel) matches or slightly exceeds MERT-v1\textsubscript{95M} on event metrics (Ev.F1@200 .763 vs.\ .756; IoU .712 vs.\ .706) at $<$2\% of its parameter cost. CultureMERT\textsubscript{95M} lags MERT-v1 across all metrics, mirroring the classification finding.

This split is notable given that sigimsae is defined by pitch contour: ornament identity is pitch-based, while localization likely draws on spectral cues such as energy envelope and timbral transitions.

\section{Design Trade-offs}\label{sec:limitations}

\textbf{Recording conditions.}
The fixed-tonic and fixed-tempo recording protocol (Section~\ref{subsec:recording}) constrains the temporal and tonal flexibility that is integral to traditional Korean vocal expression. Some singers reported that this constraint felt unnatural, which may limit the expressive freedom captured in the recordings. We treat this as a necessary trade-off for reproducible corpus-level analysis, providing a foundation upon which future work with more naturalistic recordings can build.

\textbf{Sigimsae ontology.}
The ontology prioritizes cross-genre consistency over within-genre granularity: genre-specific vocal-production techniques (e.g., mok, seong-eum) and mode-level distinctions (e.g., the descending flick in gyeongtori vs. menaritori) are absorbed into pitch-contour-based parent categories, and annotators less familiar with a given genre may default to the closest general type. This is an explicit design choice that enables systematic cross-genre comparative analysis (Section~\ref{subsec:classification}); finer-grained, genre-internal extensions remain open for future work.

\section{Applications \& Conclusions}\label{sec:conclusion}

This paper presented Gugak-VocalSet, a 24.5-hour corpus of Korean traditional vocal music with a unified, functional-position-based sigimsae ontology and approximately 68,000 audio-grounded annotations. By providing controlled-environment recordings with cross-genre annotations, the dataset enables systematic corpus-level comparative analysis of this tradition. Baseline experiments demonstrated the ontology's discriminative capacity, with domain-informed pitch representations remaining competitive against general-purpose music foundation models.

This dataset broadens the cultural scope of MIR research and lays computational groundwork for analyzing the distinctive expressive system of Korean traditional music. The annotations enable a range of downstream tasks—from analytical applications such as large-scale gugak archive analysis and singer/genre identification, to interactive applications such as automatic feedback for performance education, and to generative applications such as controlled singing voice synthesis that preserves sigimsae fidelity.

\section{Acknowledgments}
This work was carried out in collaboration with the Korea Culture Information Service Agency and the National Gugak Center, and was supported by the Ministry of Culture, Sports and Tourism through the AI Training Data Construction and Utilization Service Program for the Cultural Sector. We thank the Instrument Research Division of the National Gugak Center for curating the repertoire, and the twenty singers, the recording engineers, and the fourteen annotators whose work made this corpus possible. We also gratefully acknowledge Gwon Soo Lee of the Korea Culture Information Service Agency for his dedicated support and contributions to the successful execution of this project.
\bibliography{ISMIRtemplate}

@article{panteli2018review,
  title={A review of manual and computational approaches for the study of world music corpora},
  author={Panteli, Maria and Benetos, Emmanouil and Dixon, Simon},
  journal={Journal of New Music Research},
  volume={47},
  number={2},
  pages={176--189},
  year={2018},
  publisher={Taylor \& Francis}
}

@article{li2023mert,
  title={Mert: Acoustic music understanding model with large-scale self-supervised training},
  author={Li, Yizhi and Yuan, Ruibin and Zhang, Ge and Ma, Yinghao and Chen, Xingran and Yin, Hanzhi and Xiao, Chenghao and Lin, Chenghua and Ragni, Anton and Benetos, Emmanouil et al},
  journal={The Twelfth International
Conference on Learning Representations, ICLR 2024,
Vienna, Austria, May 7-11},
  year={2024}
}

@article{kim2009folk_sigimsae,
  title={An Examination of the Types of Sigimsae in Korean Folk Songs},
  author={Junghi Kim},
  journal={Studies In Korean Music},
  pages={49--80},
  year={2009}
}

@article{Lee1999Sigimsae,
  title={An Essay on the Method of Shigimsae Studies},
  author={Bohyeong Lee},
  journal={音樂論壇},
  volume={13},
  pages={pp.1-8},
  year={1999}
}

@article{yamamoto2022analysis,
  title={Analysis and detection of singing techniques in repertoires of J-POP solo singers},
  author={Yamamoto, Yuya and Nam, Juhan and Terasawa, Hiroko},
  journal={Proceedings of the 23rd International Society for Music Information Retrieval Conference},
  year={2022}
}

@article{Myunghyeon2014NGC,
  title={Archiving the Korean Traditional Music and Dance in National Gugak Center - An emphasis on the current status of collecting the data -},
  author={Hyeon Myeong and Hyeonok Kim},
  journal={Journal of Korean Traditional Performing Arts},
  volume={3},
  pages={7--26},
  year={2014}
}

@article{han2023finding,
  title={Finding tori: Self-supervised learning for analyzing korean folk song},
  author={Han, Danbinaerin and Repetto, Rafael Caro and Jeong, Dasaem},
  journal={Proceedings of the 24th International Society for Music Information Retrieval Conference (ISMIR)},
  year={2023}
}

@article{kumar2025recognizing,
  title={Recognizing ornaments in vocal indian art music with active annotation},
  author={Kumar, Sumit and Singh, Parampreet and Arora, Vipul},
  journal={IEEE Transactions on Audio, Speech and Language Processing},
  volume={34},
  pages={220--229},
  year={2025},
  publisher={IEEE}
}

@inproceedings{serra2014compmusic,
  title={Creating research corpora for the computational study of music: The case of the CompMusic project},
  author={Serra, Xavier},
  booktitle={AES 53rd International Conference: Semantic Audio; 2014 Jan 27-29; London, UK. New York: Audio Engineering Society; 2014. Article number 1-1 [9 p.].},
  year={2014},
  organization={Audio Engineering Society}
}

@article{srinivasamurthy2021saraga,
  title={Saraga: Open datasets for research on indian art music},
  author={Srinivasamurthy, Ajay and Gulati, Sankalp and Repetto, Rafael Caro and Serra, Xavier},
  journal={Empirical Musicology Review},
  volume={16},
  number={1},
  pages={85--98},
  year={2021}
}

@article{papaioannou2022dataset,
  title={A dataset for greek traditional and folk music: Lyra},
  author={Papaioannou, Charilaos and Valiantzas, Ioannis and Giannakopoulos, Theodoros and Kaliakatsos-Papakostas, Maximos and Potamianos, Alexandros},
  journal={Proceedings of the 23rd International Society for Music Information Retrieval Conference (ISMIR)},
  year={2022}
}

@article{sanidha2025ismir,
  title={Sanidha: A studio quality multi-modal dataset for carnatic music},
  author={Krishnan, Venkatakrishnan Vaidyanathapuram and Alben, Noel and Nair, Anish and Condit-Schultz, Nathaniel},
  journal={Proceedings of the 26th International Society for Music Information Retrieval Conference (ISMIR)},
  year={2025}
}

@inproceedings{irma2025dlfm,
  title={The IRMA Dataset: A Structured Audio--MIDI Corpus for Iranian Classical Music},
  author={Shafiei, Sepideh and Hakam, Shapour},
  booktitle={Proceedings of the 12th International Conference on Digital Libraries for Musicology},
  pages={36--43},
  year={2025}
}

@inproceedings{yamamoto2023primadnn,
  title={PrimaDNN’: A characteristics-aware DNN customization for singing technique detection},
  author={Yamamoto, Yuya and Nam, Juhan and Terasawa, Hiroko},
  booktitle={2023 31st European Signal Processing Conference (EUSIPCO)},
  pages={406--410},
  year={2023},
  organization={IEEE}
}

@misc{nia2022gugak,
  title        = {{Korean Traditional Music Audio and Score Dataset}},
  author       = {{National Information Society Agency}},
  year         = {2022},
  howpublished = {AI-Hub},
  note         = {Dataset No. 71470. Access restricted to Korean nationals},
  url          = {https://aihub.or.kr/aihubdata/data/view.do?dataSetSn=71470}
}

@article{zhang2024gtsinger,
  title={Gtsinger: A global multi-technique singing corpus with realistic music scores for all singing tasks},
  author={Zhang, Yu and Pan, Changhao and Guo, Wenxiang and Li, Ruiqi and Zhu, Zhiyuan and Wang, Jialei and Xu, Wenhao and Lu, Jingyu and Hong, Zhiqing and Wang, Chuxin and others},
  journal={Advances in Neural Information Processing Systems},
  volume={37},
  pages={1117--1140},
  year={2024}
}

@inproceedings{wilkins2018vocalset,
  title={VocalSet: A Singing Voice Dataset.},
  author={Wilkins, Julia and Seetharaman, Prem and Wahl, Alison and Pardo, Bryan},
  booktitle={Proceedings of the 19th International Society for Music Information Retrieval Conference (ISMIR)},
  pages={468--474},
  year={2018}
}

@inproceedings{nuttall2024svaraforms,
  title={Svara-forms and coarticulation in {Carnatic} music: an investigation using deep clustering},
  author={Nuttall, Thomas and Serra, Xavier and Pearson, Lara},
  booktitle={Proceedings of the 11th International Conference on Digital Libraries for Musicology (DLfM)},
  year={2024}
}

@inproceedings{kanatas2025culturemert,
  title={CultureMERT: Continual Pre-Training for Cross-Cultural Music Representation Learning},
  author={Kanatas, Angelos-Nikolaos and Papaioannou, Charilaos and Potamianos, Alexandros},
  booktitle={Proceedings of the 26th International Society for Music Information Retrieval Conference (ISMIR)},
  year={2025}
}

@book{kim2015gugak,
  author    = {Youngwoon Kim},
  title     = {Introduction to Korean Traditional Music},
  address   = {Seoul, Korea},
  publisher = {Eumaksekye},
  year      = {2015}
}

@article{bohyung2012jo,
  title={Musical Characteristics of Mode(선법) and Tori(토리)},
  author={Bohyeong Lee},
  journal={Studies In Korean Music},
  pages={245--271},
  year={2012}
}

@inproceedings{serra2011multicultural,
  title={A multicultural approach in music information research},
  author={Serra, Xavier},
  organization={Proceedings of the 12th International Society for Music Information Retrieval Conference},
  booktitle={Klapuri A, Leider C, editors. 2011 October 24-28; Miami, Florida (USA)},
  year={2011}
}

@article{rao2014overview,
  title={An overview of Hindustani music in the context of computational musicology},
  author={Rao, Suvarnalata and Rao, Preeti},
  journal={Journal of new music research},
  volume={43},
  number={1},
  pages={24--33},
  year={2014},
  publisher={Taylor \& Francis}
}

@article{bozkurt2014computational,
  title={Computational analysis of Turkish makam music: Review of state-of-the-art and challenges},
  author={Bozkurt, Bar{\i}{\c{s}} and Ayangil, Ruhi and Holzapfel, Andre},
  journal={Journal of New Music Research},
  volume={43},
  number={1},
  pages={3--23},
  year={2014},
  publisher={Taylor \& Francis}
}

@inproceedings{viraraghavan2017statistical,
  title={A Statistical Analysis of Gamakas in Carnatic Music.},
  author={Viraraghavan, Venkata Subramanian and Aravind, Rangarajan and Murthy, Hema A},
  booktitle={Proceedings of the 18th International Society for Music Information Retrieval Conference},
  pages={243--249},
  year={2017}
}

@article{ranjani2019compact,
  title={A compact pitch and time representation for melodic contours in Indian art music},
  author={Ranjani, HG and Srinivasamurthy, Ajay and Paramashivan, Deepak and Sreenivas, Thippur V},
  journal={The Journal of the Acoustical Society of America},
  volume={145},
  number={1},
  pages={597--603},
  year={2019},
  publisher={AIP Publishing}
}

@article{kwon2008TM,
  title={Archives of Korean Traditional Music Network},
  author={Dohee Kwon},
  journal={Journal of National Gugak Center},
  volume={18},
  pages={5--36},
  year={2008}
}

@article{yeri2017archive,
  title={The Preservation and Utilization of Korean Traditional Music Records -Focusing on the records housed at Kugak Archive},
  author={Yeri Noh},
  journal={The Society For Korean Historico-Musicology},
  number={59},
  pages={43--82},
  year={2017}
}

@article{injong2023gugak,
  title={Current Status and Tasks of Artificial Intelligence Technology Utilization in the Filed of Korean Traditional Music},
  author={Injong Jang and Byung-O Kim},
  journal={Journal of the Asian Music Research Institute},
  volume={54},
  pages={251--275},
  year={2023},
  publisher={Seoul National University, Music College, Asian Music Research Institute}
}

@inproceedings{gong2016pitch,
  title={Pitch contour segmentation for computer-aided Jingju singing training},
  author={Gong, Rong and Yang, Yile and Serra, Xavier},
  booktitle={Proceedings of the 13th Sound and Music Computing Conference},
  pages={172--78},
  year={2016}
}

@article{rosenzweig2020erkomaishvili,
  title={Erkomaishvili Dataset: A curated corpus of traditional Georgian vocal music for computational musicology},
  author={Rosenzweig, Sebastian and Scherbaum, Frank and Shugliashvili, David and Arifi-M{\"u}ller, Vlora and M{\"u}ller, Meinard},
  journal={Transactions of the International Society for Music Information Retrieval},
  volume={3},
  number={1},
  year={2020}
}

@inproceedings{lea2017temporal,
  title={Temporal convolutional networks for action segmentation and detection},
  author={Lea, Colin and Flynn, Michael D and Vidal, Rene and Reiter, Austin and Hager, Gregory D},
  booktitle={proceedings of the IEEE Conference on Computer Vision and Pattern Recognition},
  pages={156--165},
  year={2017}
}

@inproceedings{wei2023rmvpe,
  title={RMVPE: A Robust Model for Vocal Pitch Estimation in Polyphonic Music},
  author={Wei, Haojie and Cao, Xueke and Dan, Tangpeng and Chen, Yueguo},
  booktitle={Proceedings of the Interspeech 2023},
  pages={5421--5425},
  year={2023}
}

@article{zhou2025layer,
  title={Layer-wise investigation of large-scale self-supervised music representation models},
  author={Zhou, Yizhi and Zhu, Haina and Chen, Hangting},
  journal={arXiv preprint arXiv:2505.16306},
  year={2025}
}

@inproceedings{vocalnotes2024,
  title={The vocalnotes dataset},
  author={Proutskova, Polina and McBride, John and Ozaki, Yuto and Chiba, Gakuto and Li, Yukun and Yu, Zhaoxin and Yue, Wei and Crowdus, Miranda and Zuckerberg, Gabriel and Velichkina, Olga and others},
  booktitle={International Society for Music Information Retrieval Conference, Late-Breaking/Demo},
  year={2023}
}

@inproceedings{gong2021chmusic,
  title={Chmusic: a traditional Chinese music dataset for evaluation of instrument recognition},
  author={Gong, Xia and Zhu, Yuxiang and Zhu, Haidi and Wei, Haoran},
  booktitle={Proceedings of the 4th international conference on big data technologies},
  pages={184--189},
  year={2021}
}

@inproceedings{bhake2025melodic,
  title={Melodic and metrical elements of expressiveness in {H}industani vocal music},
  author={Bhake, Yash and Anand, Ankit and Rao, Preeti},
  booktitle={Proceedings of the 26th International Society for Music Information Retrieval Conference (ISMIR)},
  year={2025}
}

\end{document}